\documentclass[a4paper,11pt]{article}                                         %% For arXiv
\usepackage{amsmath,amssymb,mathptmx}
\numberwithin{equation}{section}                                              %% For arXiv
\newtheorem{theorem}{Theorem}[section]                                        %% For arXiv
\newtheorem{proposition}[theorem]{Proposition}                                %% For arXiv
\newtheorem{lemma}[theorem]{Lemma}                                            %% For arXiv
\newtheorem{corollary}[theorem]{Corollary}                                    %% For arXiv
\newtheorem{example}[theorem]{Example}                                        %% For arXiv
\newtheorem{conjecture}[theorem]{Conjecture}                                  %% For arXiv
\newtheorem{definition}[theorem]{Definition}                                  %% For arXiv
\newtheorem{remark}[theorem]{Remark}                                          %% For arXiv
\newenvironment{proof}{\noindent {\it Proof.} }{}                             %% For arXiv

\newcommand{\nn}{\nonumber}
\newcommand{\pp}{\partial}
\newcommand{\px}{\partial_x}
\newcommand{\ee}{\epsilon}
\newcommand{\ma}{\mathcal{A}}
\newcommand{\mb}{\mathcal{B}}
\newcommand{\me}{\mathcal{E}}
\newcommand{\mr}{\mathcal{R}}
\newcommand{\cp}{\mathcal{P}}
\newcommand{\mg}{\mathfrak{g}}
\newcommand{\lm}{\lambda}
\newcommand{\ut}{\tilde{u}}
\newcommand{\al}{\alpha}
\newcommand{\dl}{\delta}
\newcommand{\ad}{\mathrm{ad}}

\begin{document}
\title{On Quasitriviality and Integrability of a Class of Scalar Evolutionary PDEs}

\author{{Si-Qi Liu \quad Youjin Zhang} \\                                     %% For arXiv
{\small Department of Mathematical Sciences,                                  %% For arXiv
Tsinghua University Beijing 100084, P.R. China} \\                            %% For arXiv
{\small Email: lsq99@mails.tsinghua.edu.cn, youjin@mail.tsinghua.edu.cn}}     %% For arXiv

\date{}
\maketitle

\begin{abstract}
For certain class of perturbations of the equation $u_t=f(u) u_x$,
we prove the existence of change of coordinates, called quasi-Miura
transformations, that reduce these perturbed  equations to the
unperturbed ones. As an application, we propose a criterion for the
integrability of these equations.
\end{abstract}

\section{Introduction}

The notion of quasitriviality for certain class of evolutionary PDEs
was introduced in \cite{DZ01}. Let us first explain this notation by looking at the example of the Kortweg-de Vries
(KdV) equation
\begin{equation}\label{kdv}
u_t=u\,u_x+\frac{\ee^2}{12}\,u_{xxx}
\end{equation}
where $u=u(x,t)$, and $\ee$ is the dispersion parameter. By the
definition of \cite{DZ01} this equation is quasitrivial, it means
that we can perform a change of the dependent variable
\begin{equation}\label{mu-kdv-a}
v=u+ \frac{\epsilon^2}{24} \px^2 \left[ -\log u_x
+\frac{\epsilon^2}{240}  \left( 5\,\frac{u^{(4)}}{ u_x^2}
- 9\,\frac{ u_{xx}\, u_{xxx}}{ u_x^3}
+4\,\frac{u_{xx}^3}{ u_x^4}\right) + O(\epsilon^4)\right].
\end{equation}
such that in the new dependent variable the KdV equation is formally reduced to
the dispersionless equation
\begin{equation}\label{fin-1}
v_t=v\,v_x.
\end{equation}
Here the transformation (\ref{mu-kdv-a})
is understood as a formal power series of $\ee$, it has the inverse
\begin{equation}
u=v + \frac{\epsilon^2}{24} \px^2 \left[\log v_x
+\frac{\epsilon^2}{240}  \left( 5\,\frac{v^{(4)}}{ v_x^2}
- 21\,\frac{ v_{xx}\, v_{xxx}}{ v_x^3}
+16\,\frac{v_{xx}^3}{ v_x^4}\right) + O(\epsilon^4)\right].\label{mu-kdv}
\end{equation}

This property of the KdV equation was first observed in \cite{BGKI} (see also \cite{DZ01}). It
corresponds to the
genus expansion formula for the free energy of the 2d topological gravity \cite{DW,EYY,IZ}, and
is one of the important links between topological field theory
and integrable hierarchies \cite{DZ98,DZ01,EYY,K,W}.
In \cite{DZ01,DLZ}
it was proved that such a property of quasitriviality is also shared by a wide class of bihamiltonian
integrable hierarchies, and plays an important role in the study of the problem
of classification of bihamiltonian integrable hierarchies. In a very recent paper \cite{Du-05}, Dubrovin showed that
at the approximation up to $\ee^4$
any perturbation of the equation (\ref{fin-1}) which possesses a Hamiltonian structure is quasitrivial.

In the present paper we consider, without the assumption of Hamiltonian structures,
the property of quasitriviality for a
class of scalar generalized evolutionary PDEs of the form
\begin{eqnarray}
&&u_{t}=f(u) u_x+\ee(f_{1}(u) u_{xx}+f_{2}(u) u_x^2)\nn\\
&&\quad +\ee^2(f_{3}(u) u_{xxx}+f_4(u) u_{x} u_{xx}+f_5(u) u_x^3)+
\cdots,\quad f'(u)\ne 0.\label{zh-0}
\end{eqnarray}
Here
the r.h.s. of the equation is a power series of the parameter $\ee$, the coefficients
of $\ee^k$ are graded homogeneous polynomials of degree $k+1$ of the variables
$u_x, u_{xx},\dots$ with  $\deg \px^k u=k$, and
the coefficients of these polynomials are assumed to be smooth functions of $u$. Note that
when the power series of $\ee$ is a polynomial, (\ref{zh-0}) becomes a usual evolutionary PDE.

We are going to prove the quasitriviality of the equation (\ref{zh-0}) according to the following definition
of \cite{DZ01}:
\begin{definition}
The generalized evolutionary PDE (\ref{zh-0})
is called quasitrivial if there exists a quasi-Miura transformation of the form
\begin{equation}\label{zh-2}
u=v+\sum_{k\ge 1}\ee^k F_k(v,v_x,\dots,\px^{m_k} v)
\end{equation}
that formally reduces it to the equation
\begin{equation}\label{zh-05-10}
v_t=f(v) v_x.
\end{equation}
Here  $F_k,\ k\ge 1$ are smooth functions and $m_k$ are some positive integers.
\end{definition}
We call a quasi-Miura transformation of the form (\ref{zh-2}) that transforms the equation (\ref{zh-0})
to its leading term equation (\ref{zh-05-10}) the {\em reducing transformation} of (\ref{zh-0}).
The transformation (\ref{zh-2}) has an inverse which has a same form.
Note that in the original definition of quasitriviality given in \cite{DZ01,DLZ}, the coefficients
$F_k$ of the quasi-Miura transformation
are required to be rational in the variables $u_x, u_{xx},\dots$, here we slightly generalize this notion, see
Sec. \ref{sec-4} for the explicit description of the quasi-Miura transformations that we will
encounter in this paper.

The main motivation of this work
originates  From our attempt to generalize the classification scheme given in \cite{DZ01} for
a class of bihamiltonian evolutionary PDEs, we expect that the requirement of
bihamiltonian property can be replaced by a weaker restriction.
The first step along this line is to find a more general class of evolutionary PDEs
that possess the quasitriviality property, since this property of the equations
plays an important role in the classification scheme\cite{DZ01,DLZ,LZ}.
Another motivation of our study comes  From the application of the reducing transformation
to the perturbative study of solutions of the equation (\ref{zh-0}), see for example \cite{Du-05}
in which such reducing transformations are used to study the critical behavior of solutions of
the perturbed equations.

A direct consequence of the quasitriviality property is the existence of infinitely many flows of the form
\begin{equation}
u_s=h(u) u_x+\sum_{k\ge 2} \ee^{k-1} W_k(u,u_x,\dots, \pp^{n_k} u)
\end{equation}
that commute with the flow (\ref{zh-0}), here $h(u)$ is an arbitrary
smooth function. By imposing the conditions of polynomial dependence
of the functions $W_k$ on the variables $u_x,\dots, \pp^{n_k} u$, we
propose a criterion for the formal integrability of the equation
(\ref{zh-0}).

The plan of the paper is as follows: In Sec. \ref{sec-2} and Sec. \ref{sec-3} we introduce some basic notations
including our Definition \ref{zh7} of formal integrability for the equations of the form (\ref{zh-0}),
and prove an important property
of the formally integrable equations in Theorem \ref{thm-9}.
In Sec. \ref{sec-4} we prove the quasitriviality of the equations of the
form (\ref{zh-0}), the main results are summarized in Theorem \ref{zh-thm1}, \ref{main-cor-zh}.
In Sec. \ref{sec-5} we describe a criterion of formal integrability,
and illustrate it by some examples. Finally, in the conclusion we discuss the generalization of the
quasitriviality property to systems of evolutionary PDEs.

\section{Miura type transformations}\label{sec-2}

We first introduce some notations that will be used in this paper, see \cite{DZ01} for more detailed expositions
for such notations.
Let $u(x)$ be a smooth function of a real variable $x$ and denote $u_0=u(x),\ u_s=\px^s\,u(x),\ s\ge 1$.
We define the ring $\mr$ of differential polynomials of $u(x)$ as
\begin{equation}\mr=C^\infty(u_0)[u_1,u_2,\dots].\end{equation}
It is a graded ring with $\deg u_i=i,\ i\ge 1,\ \deg h(u_0)=0 \ \mbox{for any smooth function}\ h$.
We denote by $\ma$ the ring of formal power series of an indeterminate $\ee$ of the form
\begin{equation}f=\sum_{i\ge 0} f_i(u_0,\dots,u_i) \ee^i\end{equation}
where $f_i\in \mr$ are homogeneous differential polynomials of degree $i$.

The derivations of $\ma$ form a Lie algebra
\begin{equation}\mg=\{\hat{X}=\sum_{s\ge 0} X_s \frac{\pp}{\pp u_s}|X_s \in \ma\}\end{equation}
with Lie bracket
\begin{equation}[\sum_{s\ge 0} X_s\, \frac{\pp}{\pp u_s},\sum_{s\ge 0} Y_s\, \frac{\pp}{\pp u_s}]=
\sum_{s\ge 0} Z_s \frac{\pp}{\pp u_s},\mbox{ where }
Z_s=\sum_{t \ge 0}\left(X_t \frac{\pp Y_s}{\pp u_t}-Y_t \frac{\pp X_s}{\pp u_t}\right).\end{equation}
We regard the ring $\ma$ as the coordinate ring of an infinite dimensional manifold, and the Lie algebra $\mg$ as
the Lie algebra of vector fields of this manifold.

Introduce the differential operator $\px\in \mg$  by
\begin{equation}\px=\sum_{s\ge 0} u_{s+1} \frac{\pp}{\pp u_s}.\end{equation}
An element $\hat{X}\in\mg$ is called an evolutionary vector field if $[\hat{X},\px]=0$, which implies
\begin{equation}
\hat{X}=\sum_{s\ge 0} \left(\px^s X \right)\frac{\pp}{\pp u_s} \ \mbox{ for certain}\ X \in \ma.
\end{equation}
The function $X$ is called the component of the evolutionary vector field $\hat{X}$.
Denote
\begin{equation}\me=\{{\mbox{the vector space of evolutionary vector fields}}\}.\end{equation}
Then we readily have the following proposition:
\begin{proposition}
The vector space $\me$ is a Lie subalgebra of $\mg$ with the center $\{a\, \px|a\in {\mathbb R}\}$,
and its Lie bracket can be expressed as
\begin{equation}
[\hat{X},\hat{Y}]=\hat{Z}, \mathrm{\ where\ } Z=\hat{X}(Y)-\hat{Y}(X),\ \forall\ \hat{X},\hat{Y} \in \me.
\end{equation}
\end{proposition}

Since the map $X \mapsto \hat{X}$ is a bijection between $\ma$ and $\me$, we can pullback the Lie bracket
of $\me$ to $\ma$. Then we obtain a Lie bracket on $\ma$
\begin{equation}\label{bra-2}
[X,Y]=\hat{X}(Y)-\hat{Y}(X)=\sum_{s \ge 0}\left(\left(\px^s X\right)\,\frac{\pp Y}{\pp u_s}-\left(\px^s Y\right)\,
\frac{\pp X}{\pp u_s}\right).
\end{equation}
Henceforth, we will also call an element $X$ of ${\mathcal A}$ a vector field.

The subalgebra $\mb$ of the Lie algebra $(\ma,[\cdot,\cdot])$ defined by
\begin{equation}\mb=\{X=\sum_{i\ge 1} f_i(u_0,\dots,u_i) \ee^i\in\ma|\ \deg f_i=i\}\end{equation}
corresponds to the class of evolutionary PDEs that we study in this
paper. Namely, if we express $X$ in the form
\begin{eqnarray}
&&X=\ee\,f(u)\,u_x+\ee^2(f_{1}(u)\,u_{xx}+f_{2}(u)\,u_x^2)\nn\\
&&\qquad+\ee^3(f_{3}(u)\,u_{xxx}+f_4(u)\,u_{x}\,u_{xx}+f_5(u)\,u_x^3)+\cdots,\label{zh4-b}
\end{eqnarray}
and assume that it satisfies the condition $f'(u)\ne 0$, then
\begin{equation}\label{zh4} u_t={\ee^{-1}} X \end{equation}
is just the equation (\ref{zh-0}).
A vector field  $X\in\mb$ that satisfies the above condition will be called generic.
We call $\ee f(u) u_x$ the leading term of the vector field $X$, and the equation
(\ref{zh-05-10})
the leading term equation of (\ref{zh4}).

In the next section, we will consider the normal forms of the generalized evolutionary PDEs
of the form (\ref{zh-0})
under the Miura type transformations.

\begin{definition}[\cite{DZ01}]
A Miura type transformation is a transformation on the ring $\ma$ which has the form
\begin{equation}
u \mapsto \ut=X_{0}(u)+\ee\,X_{1}(u)\,u_1+\ee^2\left(X_{2}(u)\,u_2+X_{3}(u)\,u_1^2\right)+\cdots,
\ \frac{\pp X_{0}(u)}{\pp u}\ne 0.
\end{equation}
\end{definition}
A Miura type transformation is an automorphism of the ring $\ma$, and it induces an automorphism of the Lie algebra $\mb$.
It is easy to see that all such transformations form a group which is called the Miura group \cite{DZ01}.
It is the direct product of two subgroups.
The first subgroup is the diffeomorphism group of $\mathbb{R}$
\begin{equation} u\mapsto\ut=f(u),\mathrm{\ where\ }\frac{\pp f}{\pp u}\ne 0.
\end{equation}
The second subgroup is formed by the Miura type transformation with $X_{0}(u)=u$. One can prove that any Miura
type transformation with $X_{0}(u)=u$ can be expressed as
\begin{equation}
g:\ u\mapsto\ut=e^{\hat{Y}} u=u+\hat{Y}(u)+\frac12\hat{Y}(\hat{Y}(u))+\frac16\hat{Y}(\hat{Y}(\hat{Y}(u)))+\cdots
\end{equation}
where $Y=\ee\,Y_{1}(u) u_1+ \ee^2\left(Y_{2}(u)\,u_2+Y_{3}(u)\,u_1^2\right)+\cdots\in \mb$.
The automorphism of $\mb$ induced by this Miura type transformation $g$ is given by
\begin{equation}
X\mapsto g(X)=e^{-\ad_Y}X=X-[Y,X]+\frac12[Y,[Y,X]]-\frac16[Y,[Y,[Y,X]]]+\cdots.
\end{equation}
Here the vector field $g(X)$ is obtained by first expressing the vector field $X$ in the
new coordinate $\tilde u$, then
redenote $\tilde u$ by $u$.
We call the transformations of these two subgroups the Miura type transformations of the first
and second kind respectively.

\section{Formal symmetries and integrability}\label{sec-3}

In this section, we first give the definition of formal symmetries for a generalized evolutionary
PDE of the form (\ref{zh-0}) (or equivalently, (\ref{zh4}) for a generic vector field $X\in\mb$),
then we study their properties and based on which we
introduce the notion of formal integrability. We show that the formal integrability of a
equation of the form (\ref{zh-0}) is equivalent to
the existence of a unique reduced form of the equation under Miura type transformations.

\begin{definition}\label{zh6}
Given a generic vector field $X\in \mb$ of the form (\ref{zh4-b}), a formal symmetry of the equation (\ref{zh-0})
is a flow of the form
\begin{equation}\label{sym-zh1}
\ee \pp_s u=Y,\quad Y\in\mb
\end{equation}
which commutes with (\ref{zh-0}). We will also call $Y$ a formal symmetry of the vector field $X$.
\end{definition}
This definition is adopted  From the usual definition of generalized infinitesimal symmetries
of an evolutionary PDE. It differs  From the usual one at the following two points.
Firstly, even when (\ref{zh-0}) is a usual evolutionary PDE whose right hand side is truncated,
i.e. a polynomial of the finite number of variables $u_1,\dots, u_m$ for certain positive integer $m$,
its formal symmetries are not necessarily truncated. While in the usual definition an infinitesimal symmetry
depend only on finite number of the variables $u_i,\ i\ge 1$. Secondly, the right hand side of a formal
symmetry (\ref{sym-zh1})
is required to have the same form (\ref{zh4-b}) as of the vector field $X$
(except the condition $f'(u)\ne 0$). These features of the formal symmetries will play a crucial
role in our discussion of formal integrability of the equations of the form (\ref{zh-0}).

Before the exposition of properties of the formal symmetries, let us
first do some preparations. Let $\cp$ be the set of all ordered sequences of integers $\lm=(\lm_1,\lm_2,\dots)$
with the properties
\begin{equation}
\lm_1\ge \lm_2\ge\dots\ge \lm_m\ge 0,\quad |\lm|=\sum\limits_{i\ge1}\lm_i<\infty.
\end{equation}
We call $|\lm|$ the degree of $\lm$,
and denote by $\cp^d$ the subset of degree $d$ elements of $\cp$.
Each $\lm\in\cp$ corresponds to a unique monomial
\begin{equation}
u_\lm=\prod_{i\ge 1}u_{\lm_i}\in \mr.
\end{equation}
Given two elements $\lm,\mu \in \cp^d$, we say $\lm > \mu$ (resp. $\lm < \mu$) if the first non-zero entry
of the sequence $(\lm_1-\mu_1,\lm_2-\mu_2,\lm_3-\mu_3,\dots)$ is greater (resp. less) than $0$.
By using this ordering of $\cp^d$, we can define in a natural way the highest order term of a
homogeneous differential polynomial
$X\in \mr$ of degree $d$. For example, the highest order term of
\begin{equation}
X=X_1(u)u_4 u_1+X_2(u)u_3 u_2+X_3(u)u_2 u_2 u_1
\end{equation}
is $X_1(u)u_4 u_1$, and we denote the sum of all other terms in $X$ by $l.o.t$.
\begin{lemma}\label{thm-3}
Let $X\in\mr$ be homogeneous of degree $d$ of the form
\begin{equation}\label{xx-3}
X=f(u)\prod_{k=1}^m \left(u_k\right)^{\al_k}+l.o.t,
\end{equation}
then we have
\begin{equation}
[u\,u_1,X]=\left(\sum_{k=2}^{m}\al_k+d-1\right)f(u)\,u_1\prod_{k=1}^m \left(u_k\right)^{\al_k}+l.o.t .
\end{equation}
\end{lemma}
\begin{proof}
After the substitution of (\ref{xx-3}) into (\ref{bra-2}), we can prove the lemma by a straightforward computation.
\hfill $\Box$ \end{proof}

\begin{lemma}\label{thm-4}
Assume $X\in \mr$ be given as in the above lemma.
Then $[u\,u_1,X]=0$ if and only if $d=1$, i.e. $X$ has the form $X=f(u)u_1$.
\end{lemma}
\begin{proof}
Let $X$ have the form (\ref{xx-3}) and satisfies $[u\,u_1,X]=0$, then  From Lemma
\ref{thm-3} we obtain $\al_1=1, \al_2=\cdots=\al_m=0$, so $X=f(u) u_1$. Conversely, it is obvious that
the identity $[u\,u_1,X]=0$ holds true for any vector field $X$ of the form $f(u)u_1$. The
lemma is proved.
\hfill $\Box$ \end{proof}

\begin{lemma}\label{thm-5}
For a homogeneous differential polynomial $Y\in\mr$ of degree $d>2$,
if the highest order term of $Y$ has a factor $u_1$,
then we can find a homogeneous differential polynomial $X\in\mr$ of degree $d-1$ such that $[u_0\,u_1,X]$ has the same
highest order term as that of $Y$.
\end{lemma}
\begin{proof}
When $d\ge 2$, we always have $\left(\sum\limits_{k=2}^{m}\al_k+d-1\right)\ne0$. So the lemma immediately
follows  From Lemma \ref{thm-3}.
\hfill $\Box$ \end{proof}

\begin{theorem}\label{thm-7}
Let $Y\in\mb$ be a symmetry of a generic vector field $X\in\mb$ and
$Y\ne a u_x, \ a\in{\mathbb R}$.
Then $Y$ is also generic
and is determined by its leading terms.
\end{theorem}
\begin{proof}
We only need to prove that symmetries are determined by their leading terms.
Let $Y,Z \in \mb$ be two symmetries of the generic vector field $X\in\mb$ and have the same
leading terms. Then  From the identity
\begin{equation}\label{zh2}
[X, Y-Z]=[X, Y]-[X,Z]=0
\end{equation}
it follows that $W=Y-Z$ is also a symmetry of $X$.
Sine $Y, Z$ have the same leading terms, the vector field $W$ has the following expression
\begin{equation}
W
=\sum_{d \ge 2}\ee^d\,W_d(u,u_1,\dots,u_d).
\end{equation}
Denote
\begin{equation}
X=\sum_{d\ge 1} \ee^d X_d(u,\dots, u_d)
\end{equation}
Then  From (\ref{zh2}) we know that the vector field $\ee^2 W_2$ is a symmetry of the vector field
$\ee X_{1}(u,u_1)$.
According to Lemma \ref{thm-4}, symmetries of $\ee X_{1}(u,u_1)$ must be of degree $1$, so we have $W_2=0$.
Similarly, we can prove that all the $W_m$ vanish and consequently, $Y=Z$.
The theorem is proved.
\hfill $\Box$ \end{proof}

\begin{corollary}\label{thm-8}
Let $X\in\mb$ be generic, $Y_1,Y_2$ are two symmetries of $X$, then $[Y_1,Y_2]=0$.
\end{corollary}
\begin{proof}
By using the Jacobi identity we know that $[Y_1,Y_2]$ is also a symmetry of $X$. But the leading terms
of $[Y_1,Y_2]$ vanish, so  From Theorem \ref{thm-7} it follows that $[Y_1,Y_2]=0$.
\hfill $\Box$ \end{proof}

\begin{remark}
Let us denote $X \sim Y$ if $[X,Y]=0$, then the above corollary shows that $\sim$ is an equivalence relation on
the set of generic vector fields of $\mb$.
\end{remark}

\begin{theorem}\label{thm-6}
For any generic $X\in\mb$, there exists a Miura type transformation $g$ such that
\begin{equation}\label{normal-2}
g(X)=\ee\,u\,u_1+\ee^2\left(f_{(2)}u_2+f_{(1,1)}u_1^2\right)+\sum_{d \ge 3}\ee^d \left(\sum_{\lm \in \cp_1^d}
f_{\lm}(u)u_{\lm}\right),
\end{equation}
where $\cp_1^d$ is the set of partitions of degree $d$ whose nonzero entries are greater than $1$.
\end{theorem}
\begin{proof}
Let $\ee f(u) u_1$ be the leading term of the vector field $X$.
We can use a Miura type transformation of the first kind $\ut=f(u)$ to transform $X$ to the following form
\begin{equation}
g(X)=\ee\,u\,u_1+\sum_{d \ge 2}\ee^d \left(\sum_{\lm \in \cp^d}f_{\lm}(u)u_{\lm}\right).
\end{equation}
Then by using Lemma \ref{thm-5}, we can find a series of Miura type transformations of the second kind to
eliminate the terms with factor $u_1$ step by step (except for the terms of degree $2$). The theorem is proved.
\hfill $\Box$ \end{proof}
The expression (\ref{normal-2}) is called a reduced form of the vector field $X$. Note that
such a reduced form may not be unique.

\begin{definition}\label{zh7}
A generalized evolutionary PDE of the form (\ref{zh-0}) that corresponds to a
generic vector field $X\in\mb$ of the form (\ref{zh4-b})
is called formally integrable, if any vector field of the form
$Y=\ee\,h(u) u_1\in \mb$ can be extended to a symmetry
\begin{equation}
\tilde{Y}=\ee h(u)\,u_1+\sum_{k\ge2}\ee^k Y_k\in\mb.
\end{equation}
of this equation.
\end{definition}
By definition, a formally integrable equation of the form
(\ref{zh-0}) must have infinitely many formal symmetries. Thus this
definition is an adaption of the commonly used one of
integrability  From the point of view of symmetries in soliton
theory, see for example \cite{Fokas1,Fokas2,MSY,MSS,SW1,SM} and references therein,
to the present class of generalized evolutionary PDEs. In \cite{Fokas1,Fokas2}, it was
conjectured by Fokas
that the existence of a single time independent non-Lie point symmetry implies the existence of infinitely many
for a scalar evolutionary PDE. We will reformulate this conjecture for our class of equations and their
formal symmetries in Sec.\,\ref{sec-5}.
One of the important contributions
of the approach of Mikhailov, Shabat and their collaborators \cite{MSY,MSS,SM}
is the formulation of certain necessary conditions of the existence of a higher order
symmetry for a system of evolutionary PDEs. These conditions are expressed in terms of the so called
canonical conservation laws, they yield an effective algorithm of testing the existence of
a higher order symmetry for certain class of evolutionary PDEs of lower order. In \cite{SW1},
Sanders and Wang gave a symbolic algorithm of checking the existence of symmetries for certain
class of scalar
evolutionary PDE, such equations depend polynomially on the dependent variable and its $x$-derivatives,
and satisfy certain homogeneity conditions. We will give in Sec.\,\ref{sec-5} an alternative
way of checking the formal integrability of the generalized evolutionary PDEs, our approach is base
on the quasitriviality of such equations which will be proved in the next section.

The following theorem relates the uniqueness of the reduced form of a vector field to the
integrability of the corresponding PDE.
\begin{theorem}\label{thm-9}
A generic vector field $X\in\mb$ of the form (\ref{zh4-b}) has a unique reduced form if and only if the
corresponding equation (\ref{zh-0}) is formally integrable.
\end{theorem}
\begin{proof}
Without the loss of generality, we assume that $X$ is already in a reduced form
that is given by the right hand side of (\ref{normal-2}).

Let us first suppose that the reduced form of $X$ is unique.
Given a vector field
$$\ee\,Y_1=\ee\,h(u)\,u_1 \in\mb.$$
Consider the following vector field
\begin{equation}
g_1(X):=e^{\ad_{\ee Y_1}} X
=\ee\,u_0\,u_1+\ee^2\left(f_{(2)}u_2+f_{(1,1)}u_1^2\right)+\cdots,
\end{equation}
where $g_1$ corresponds to the Miura type transformation
\begin{equation}
g_1:\ u\mapsto e^{-{\ee \hat{Y}_1}} u .
\end{equation}
 From the proof of Theorem \ref{thm-6}, we know that there exists a Miura transformation
\begin{equation}
g_2:\ u\mapsto e^{{\hat Z}} u,\quad Z=\ee^2 Z_2+\ee^3 Z_3+\dots\in\mb
\end{equation}
which transforms $g_1(X)$ to a reduced form. It follows  From our assumption on the uniqueness of reduced
form for the vector field $X$ that
$$
X=e^{-\ad_{{Z}}} g_1(X) .
$$
On the other hand, $e^{-\ad_{{Z}}} e^{\ad_{\ee Y_1}} X$ can be rewritten as $e^{\ad_Y} X$
with a vector field $Y$ of the form
$$
Y=\ee Y_1+\ee^2 Y_2+\cdots\in\mb.
$$
 From the equality $X=e^{\ad_Y} X$ it then follows that the vector field $X$ has a symmetry $Y$
with the leading term $\ee Y_1$. Thus by our definition the equation $\ee u_t=X$ is integrable.

Now let us assume that the equation $\ee u_t=X$ is integrable. If $X$ has another reduced form
$\tilde X$, then it must be related to $X$ by a Miura type transformation
\begin{equation}
\tilde X=e^{\ad_{Y}} X,\quad Y=\ee Y_1+\ee^2 Y_2+\dots\in\mb.
\end{equation}
 From the definition of integrability, we can find a symmetry $Z$ of the vector field $X$
that has the same leading term as the vector field $Y$, i.e.,
\begin{equation}
Z=\ee Y_1+\sum_{k\ge 2} \ee^k Z_k \in\mb.
\end{equation}
So we can express $\tilde X$ as
\begin{equation}
\tilde X=e^{\ad_{Y}} e^{-\ad_Z} X=e^{\ad_W}X.
\end{equation}
Here the vector field $W$ has the form $W=\ee^2 W_2+\cdots$. Since both $X$ and $\tilde X$
are expressed in reduced forms, by using Lemma \ref{thm-3} we know that $W=0$. Thus $\tilde X=X$ and
we proved the uniqueness of reduced form of the vector field $X$.
The theorem is proved.
\hfill $\Box$ \end{proof}

\section{Quasitriviality}\label{sec-4}

We now proceed to consider the quasitriviality of a generalized evolutionary PDE of the form (\ref{zh-0})
corresponding to a generic vector field $X$ of the form (\ref{zh4-b}).
Define a map  From the set of infinite series to itself
\begin{equation}
%R: (u_1,u_2,\dots)\mapsto (\phi^1,\phi^2,\dots)
R: (y_1,y_2,\dots)\mapsto (z_1,z_2,\dots)
\end{equation}
in the following recursive way
\begin{equation}
z_1=\frac1{y_1},\ z_n=\frac1{y_1} \sum_{k\ge 1} y_{k+1} \frac{\pp z_{n-1}}{\pp y^k},\quad
n\ge 2.
\end{equation}
When $y_k$ are given by the $k$-th order derivatives of a single variable smooth function $A$, then the above define
$z_k$ are just the $k$-th order derivatives of the
inverse function of $A$. By using this observation it is easy to see that $R$ is an involution, i.e. $R^2$ is the identity map.
\begin{lemma}\label{kerlem}
The equation $[u\,u_1,f(u,u_1,\dots)]=0$ for the unknown function $f$ has the following general solution
\begin{equation}
f=\frac1{\phi_1}c(u,\phi_2,\phi_3,\dots)\label{sol-1},
\end{equation}
where $c$ is an arbitrary smooth function and $(\phi_1,\phi_2,\dots)=R(u_1,u_2,\dots)$.
\end{lemma}
\begin{proof}
A solution of the equation $[u\,u_1,f(u,u_x,\dots)]=0$ corresponds to a flow $\frac{\pp u}{\pp s}=f$ that commutes with the
flow $\frac{\pp u}{\pp t}=u u_x$. By performing a transformation
\begin{equation}
(x,t,s, u(x,t,s))\mapsto (u,t,s, x(u,t,s)),
\end{equation}
we can rewrite these two flows into the form
\begin{equation}
\frac{\pp x}{\pp s}=g(u,x_u,x_{uu},\dots),\quad \frac{\pp x}{\pp t}=-u
\end{equation}
Here $g=-f(u,\frac{1}{x_u}, \frac1{x_u} \pp_u \frac{1}{x_u},\dots)\,\frac{\pp x}{\pp u}$. The condition of commutativity of the
flows is given by
$$\frac{\pp g}{\pp t}=0.$$
Due to the fact that
$\frac{\pp}{\pp t}\frac{\pp^k x}{\pp u^k}=\frac{\pp^k}{\pp u^k}\frac{\pp x}{\pp t}=-\frac{\pp^k u}{\pp u^k}=-\delta_{k,1}$,
we know that the general solution of the above equation has the form $g=-c(u,x_{uu},x_{uuu},\dots)$.
for certain smooth function $c$. So $f$ must have the form (\ref{sol-1}), and the lemma is proved.
\hfill $\Box$ \end{proof}

\begin{proposition}\label{mainlem}
For any given smooth function $F(u,u_1,\dots)$, the equation $[u u_1,f]=F$ for the unknown function
$f=f(u,u_1,\dots)$
has a particular solution
\begin{equation}
f=-u_1\left.\left(\int^{\phi_1} \phi_1\,F(u, \tilde{u}_1, \tilde{u}_2,\dots)
d\,\phi_1\right)\right|_{(\phi_1,\phi_2,\dots)=R(u_1,u_2,\dots)},
\end{equation}
where $(\tilde{u}_1, \tilde{u}_2, \dots)=R(\phi_1, \phi_2, \dots)$.
\end{proposition}
\begin{proof}
Assume that we have a solution of the form
$f=\frac1{\phi_1}\, c(u,\phi_1,\phi_2,\phi_3,\dots)$.
By using the above theorem we know that the function $c$ must satisfy the
following equation
\begin{equation}\label{zh5}
\frac{\pp c}{\pp \phi_1}=-\phi_1\,F(u,u_1,u_2,\dots)=-\phi_1\,F(u, \frac1{\phi_1},-\frac{\phi_2}{\phi_1^3},\dots).
\end{equation}
Here in the second equality we used the substitution
\begin{equation}
(u_1, u_2,\dots)= R(\phi_1,\phi_2,\dots)=(\frac1{\phi_1},-\frac{\phi_2}{\phi_1^3},\dots).
\end{equation}
By integrating (\ref{zh5}) with respect to $\phi_1$ and making  the substitution
$$(\phi_1, \phi_2, \dots)=R(u_1,u_2,\dots)$$
 we finish the proof of the lemma.
\hfill $\Box$ \end{proof}

Now we are ready to prove the existence of a reducing transformation for the
generalized evolutionary PDEs.
\begin{theorem}\label{zh-thm1}
Given a generalized evolutionary PDE of the form
\begin{equation}\label{zh-1}
u_t=u u_x+\ee\left(f_1(u) u_2+f_2(u) u_1^2\right)+\sum_{k\ge 3} \ee^{k-1} S_k,
\end{equation}
with $S_k\in\mr$ being homogeneous differential polynomials of degree $k$, we have

\noindent i) there exists a
unique transformation of the form
\begin{eqnarray}\label{zh-2a}
&&u\mapsto v=e^{\hat{X}} u,\\
&&X=\sum_{k\ge1}\ee^k\,u_1^{-L_k}\,
\sum_{m=0}^{M_k}Y_{k,m}(u,u_1,\dots,u_{N_k}) \left(\log{u_1}\right)^m \label{zh-2b}
\end{eqnarray}
that reduces the equation (\ref{zh-1}) to ${v}_t={v} {v}_x$.
Here $Y_{k,m}\in\mr$ are homogeneous differential polynomials of degree $L_k+k$,
$L_k,M_k,N_k$ are integers which only depend on $k$.

\noindent ii) The transformation (\ref{zh-2a}), (\ref{zh-2b}) also reduces any symmetry
\begin{equation}\label{zh-3}
u_s=h(u) u_1+\sum_{k\ge 2} \ee^{k-1} Q_k,\quad Q_k\in\mr,\ \deg Q_k=k
\end{equation}
of the equation (\ref{zh-1}) to the form $v_s=h(v) v_x$.
\end{theorem}
\begin{proof}
Let the reducing transformation takes the following form
\begin{equation}
u\mapsto v=e^{\hat{X}}(u)=u+\hat{X}(u)+\frac12\hat{X}(\hat{X}(u))+\frac16\hat{X}(\hat{X}(\hat{X}(u)))+\cdots
\end{equation}
where $X=\ee\,X_1+\ee^2\,X_2+\ee^3\,X_3+\ee^4\,X_4+\cdots$.
It will eliminate all the perturbations, so we have
\begin{equation}
e^{\ad_X}(\ee\,S_1)=\ee\,S_1+\ee^2\,S_2+\ee^3\,S_3+\ee^4\,S_4+\ee^5\,S_5+\cdots
\end{equation}
where $S_1=u u_1,\ S_2=f_1(u) u_2+f_2(u) u_1^2$.
The coefficients of $\ee^k$ give us the following equations
\begin{eqnarray*}
S_2&=&[X_1,u u_1], \\
S_3&=&[X_2,u u_1]+\frac12[X_1,[X_1,u u_1]], \\
S_4&=&[X_3,u u_1]+\frac12[X_2,[X_1,u u_1]]+\frac12[X_1,[X_2,u u_1]]+\frac16[X_1,[X_1,[X_1,u u_1]]]\nn
\end{eqnarray*}
and so on. By using Proposition \ref{mainlem}, we can solve these equations and obtain $X_1$, $X_2$, $X_3$, $X_4$, $\dots$
recursively, and the homogeneity condition on $Y_{k,m}$ guarantees the uniqueness of the solution $X_k, k\ge 1$.
Thus we proved the first part of the theorem. To prove the second result of the theorem,
let us note that after the transformation (\ref{zh-2a}), (\ref{zh-2b}) the equation
(\ref{zh-3}) is transformed to
\begin{equation}
u_s=e^{-\ad_{X}}(h(u) u_1+\sum_{k\ge 2} \ee^{k-1} Q_k)=h(u) u_x+\sum_{k\ge 2} \ee^{k-1} {\tilde Q}_k(u, u_1,\dots,u_{m_k})
\end{equation}
here for the simplicity of notations we keep to use the symbol $u$ instead of $v$, and ${\tilde Q}_k$ are homogeneous
polynomials of $\log{u_1}, u_1,\frac1{u_1}, u_2,\dots, u_{m_k}$ of degree $k$ (note that $\deg\log{u_1}=0$), where $m_k$
is a positive integer depending on $k$.
Since the transformation (\ref{zh-2a}), (\ref{zh-2b}) reduces the equation (\ref{zh-1})
to the form $u_t=u u_x$ we know that
\begin{equation}
[u u_x,h(u) u_x+\sum_{k\ge 2} \ee^{k-1} {\tilde Q}_k(u, u_1,\dots,u_k)]=0
\end{equation}
By using Lemma \ref{kerlem} and the fact that $\deg Q_k=k>1$, we see that
the functions ${\tilde Q}_k$ must vanish. The theorem is proved.
\hfill $\Box$ \end{proof}

\begin{remark}
We conjecture that the integers $L_k, M_k, N_k$ has the following expressions:
$$
L_k=3k-2,\ M_k=2\left[\frac{k-1}2\right]+\delta_{0,\left[\frac{k-1}2\right]},\ N_k=k+1.
$$
We hope that they can be proved by a careful analysis of the above procedure of
the construction of the reducing transformation.
\end{remark}

By using Theorem \ref{zh-thm1} we are readily led to the following corollary:
\begin{theorem}\label{main-cor-zh}
Any generalized evolutionary PDE of the form (\ref{zh-0}) is quasitrivial.
\end{theorem}

As we already noted in the introduction, the notion of quasi-Miura transformation that was introduced in
\cite{DZ01}(see also \cite{DLZ}) requires that the functions
$F_k(u,u_x,u_{xx},\dots)$ depend rationally on the variables $u_1, u_2,\dots$. Here we drop this
rationality condition and still call (\ref{zh-2}) a quasi-Miura transformation.
For the equation (\ref{zh-0}), when $f_2 f'-f_1 f''=0$ the reducing transformation (\ref{zh-2}) depend
rationally on the variables $u_k, k\ge 1.$

\begin{example}[KdV Equation]
The KdV equation (\ref{kdv})
has the reducing transformation (\ref{mu-kdv-a}), (\ref{mu-kdv}).
\end{example}

\begin{example}[Camassa-Holm Equation]
Let us consider another important equation in soliton theory, the
Camassa-Holm equation \cite{CH,CHH,FF}
\begin{equation}\label{ch-eq-a}
u_t-u_{xxt}+3u\,u_x-2u_x\,u_{xx}-u\,u_{xxx}=0 .
\end{equation}
Let us perform the rescaling $t \mapsto -3\ee t,\ x\mapsto \ee x$, then the above equation can be put
into the following form:
\begin{eqnarray}
&&\ee\,u_t=\left(1-\ee\,\px^2\right)^{-1}\left(\ee\,u\,u_1-\ee^3\left(\frac23u_1\,u_2+\frac13u\,u_3\right)\right) \\
&&\quad=\ee\,u\,{u_1} + \ee^3\,\left( \frac{7\,{u_1}\,{u_2}}{3} + \frac{2\,u\,{u_3}}{3} \right)+
\ee^5\,\left( \frac{23\,{u_2}\,{u_3}}{3} + \frac{11\,{u_1}\,{u_4}}{3}+\frac{2\,u\,{u_5}}{3} \right)+\cdots\nn
\end{eqnarray}
It has the reducing transformation
\begin{eqnarray}
&&v \mapsto u=v + \ee^2\,\left( \frac{7\,{v_2}}{6} - \frac{v\,{{v_2}}^2}{3\,{{v_1}}^2} + \frac{v\,{v_3}}{3\,{v_1}} \right)  +
\ee^4\,\left( \frac{6\,{{v_2}}^3}{5\,{{v_1}}^2} - \frac{202\,v\,{{v_2}}^4}{45\,{{v_1}}^4}\right.\nn\\
&&+\frac{32\,v^2\,{{v_2}}^5}{9\,{{v_1}}^6} - \frac{181\,{v_2}\,{v_3}}{90\,{v_1}}
+ \frac{398\,v\,{{v_2}}^2\,{v_3}}{45\,{{v_1}}^3} - \frac{70\,v^2\,{{v_2}}^3\,{v_3}}{9\,{{v_1}}^5} -
\frac{191\,v\,{{v_3}}^2}{90\,{{v_1}}^2}\nn\\
&& + \frac{19\,v^2\,{v_2}\,{{v_3}}^2}{6\,{{v_1}}^4} + \frac{143\,{v_4}}{72} -
\frac{133\,v\,{v_2}\,{v_4}}{45\,{{v_1}}^2}+ \frac{34\,v^2\,{{v_2}}^2\,{v_4}}{15\,{{v_1}}^4} -
\frac{73\,v^2\,{v_3}\,{v_4}}{90\,{{v_1}}^3}\nn\\
&&\left. + \frac{13\,v\,{v_5}}{18\,{v_1}} -
\frac{41\,v^2\,{v_2}\,{v_5}}{90\,{{v_1}}^3} + \frac{v^2\,{v_6}}{18\,{{v_1}}^2} \right)+\cdots .\nn
\end{eqnarray}
\end{example}

Both the KdV equation and the Camassa-Holm equation have semisimple bihamiltonian structures, their
quasitriviality have been proved in \cite{DZ01} and \cite{DLZ}.
The following example of the Burgers equation does not possess hamiltonian structures.

\begin{example}[Burgers Equation]
The Burgers equation
\begin{equation}\label{burgers-eq}
u_t=u u_x+\ee\,u_{xx}
\end{equation}
has the reducing transformation
\begin{eqnarray}
&&v \mapsto u=v + \ee\,\frac{\,{v_2}}{{v_1}} + \ee^2\,\left( \frac{2\,{{v_2}}^3}{{{v_1}}^4} - \frac{7\,{v_2}\,{v_3}}{3\,{{v_1}}^3} +
\frac{{v_4}}{2\,{{v_1}}^2} \right)
+ \ee^3\,\left( \frac{24\,{{v_2}}^5}{{{v_1}}^7}\right.\nn\\
&&\left.-\frac{46\,{{v_2}}^3\,{v_3}}{{{v_1}}^6}
+ \frac{16\,{v_2}\,{{v_3}}^2}{{{v_1}}^5}+
\frac{34\,{{v_2}}^2\,{v_4}}{3\,{{v_1}}^5} - \frac{10\,{v_3}\,{v_4}}{3\,{{v_1}}^4} - \frac{11\,{v_2}\,{v_5}}{6\,{{v_1}}^4} +
\frac{{v_6}}{6\,{{v_1}}^3} \right)\nn\\
&&  + \ee^4\,\left( \frac{568\,{{v_2}}^7}{{{v_1}}^{10}} -
\frac{4544\,{{v_2}}^5\,{v_3}}{3\,{{v_1}}^9} + \frac{1086\,{{v_2}}^3\,{{v_3}}^2}{{{v_1}}^8}-
\frac{179\,{v_2}\,{{v_3}}^3}{{{v_1}}^7} + \frac{1154\,{{v_2}}^4\,{v_4}}{3\,{{v_1}}^8}\right.\nn\\
&&\left. -
\frac{380\,{{v_2}}^2\,{v_3}\,{v_4}}{{{v_1}}^7} + \frac{221\,{{v_3}}^2\,{v_4}}{6\,{{v_1}}^6} +
\frac{26\,{v_2}\,{{v_4}}^2}{{{v_1}}^6} - \frac{70\,{{v_2}}^3\,{v_5}}{{{v_1}}^7}+
\frac{731\,{v_2}\,{v_3}\,{v_5}}{18\,{{v_1}}^6}\right.\nn\\
&&\left. - \frac{101\,{v_4}\,{v_5}}{30\,{{v_1}}^5} +
\frac{83\,{{v_2}}^2\,{v_6}}{9\,{{v_1}}^6} - \frac{13\,{v_3}\,{v_6}}{6\,{{v_1}}^5} - \frac{5\,{v_2}\,{v_7}}{6\,{{v_1}}^5} +
\frac{{v_8}}{24\,{{v_1}}^4} \right)+\cdots .\nn
\end{eqnarray}
\end{example}

\begin{example}
Our last example is a class of equations parameterized
by a smooth function $f$
\begin{equation}
u_{t_f}=\ee\,u_1 f(u+ \ee\,u_1)=\sum_{n\ge0}\ee^{n+1}\,\frac{f^{(n)}(u)}{n!}\,{u_1}^{n+1}.
\end{equation}
It's easy to verify that $\pp_{t_f}$ and $\pp_{t_g}$ commute for arbitrary smooth functions
$f$ and $g$. The reducing transformation
of this equation has an explicit form
\begin{equation}
v \mapsto u=e^{\ee \hat{X}}(v)=v+\ee\,\hat{X}(v)+\frac{\ee^2}2\,\hat{X}(\hat{X}(v))+
\frac{\ee^3}6\,\hat{X}(\hat{X}(\hat{X}(v)))+\cdots,
\end{equation}
where $X=v_1 \log(v_1)$. This equation is quite different  From those of the above three examples, because
its reducing transformation contains $\log(v_1)$, while that of the KdV, the Camassa-Holm and
the Burgers equation are rational in jet variables $v_1,v_2,\dots$.
\end{example}

\section{Testing of integrability}\label{sec-5}

Given a generalized evolutionary PDE of the form (\ref{zh-0}), let
$$
g: u\mapsto v
$$
be the quasi-Miura transformation (\ref{zh-2}) that
reduces it to the form (\ref{zh-05-10}).
For any smooth function $h(v)$, the equation (\ref{zh-05-10}) has a symmetry
\begin{equation}\label{sym-2}
{v}_s=h(v) {v}_x.
\end{equation}
Rewrite this flow in the $u$ coordinates by using the quasi-Miura transformation $g^{-1}$,
we have
\begin{equation}\label{sym-1}
\ee u_s=\ee h(u) u_x+\sum_{k\ge 2} \ee^{k} W_k(u,u_1,\dots,m_k),
\end{equation}
where in general $W_k$ are not polynomials of the variables $u_1,u_2,\dots$.

In order to compute these $W_k$ in a more direct way,  we first perform a change of the dependent variable
\begin{equation}
u\mapsto f(u)
\end{equation}
to transform the equation (\ref{zh-0}) into the form
\begin{equation}
\ee u_t=\tilde X,\quad \tilde X=\ee u u_x+\sum_{k\ge 2} \ee^{k} {\tilde X}_k(u,\dots, u_k).
\end{equation}
Consider the equation $[\tilde X, Y]=0$
for the unknown vector field $$Y=\ee\,h\circ f^{-1}(u) u_x+\sum_{k\ge 2} \ee^k {Y}_k(u,\dots,u_{n_k}).$$
By using Proposition \ref{mainlem} we can solve this equation recursively to obtain $Y$. Under certain homogeneity
condition that is similar to the one given in Theorem \ref{zh-thm1} the solution $Y$ is unique. By
performing the transformation $u\mapsto f^{-1}(u)$ the vector field $Y$ is transformed to the form
$$
\ee h(u) u_x+\sum_{k\ge 2} \ee^k {\tilde Y}_k(u,\dots,u_{n_k}).
$$
Then we have $W_k={\tilde Y}_k$.

Although the above flow (\ref{sym-1}) commutes with the flow given by the equation (\ref{zh-0}), in general it does not
meet the requirement of being a formal symmetry of (\ref{zh-0}) according to our Definition \ref{zh6}.
This is because in general the right hand side of the equation (\ref{sym-1})
does not belong to $\mb$, i.e.,
the functions $W_k(u,u_1,\dots,u_{m_k})$ are not polynomials of the variables $u_1,u_2,\dots, u_{m_k}$.

Now let us come back to our Definition \ref{zh7} of formal integrability for a
generalized evolutionary PDE of the
form (\ref{zh-0}). By using the second result of Theorem \ref{zh-thm1} and the quasitriviality of
the equation (\ref{zh-0}), we know that if a vector field $\ee h(u) u_x$ can be extended to a
formal symmetry of the equation (\ref{zh-0}), then this formal symmetry must coincide with (\ref{sym-1}).
Thus we have the following:

\vspace{1ex}
\noindent{\bf Criterion of formal integrability}. {\em
A generalized evolutionary PDE of the form (\ref{zh-0})
is formally integrable iff for any smooth function $h(u)$, the functions $W_k$ that appear in
the right hand side of (\ref{sym-1})  are homogeneous differential polynomials
of $u_1, u_2,\dots, u_k$ of degree $k$.}
\vspace{1ex}

Let us consider the formal integrability of the four equations considered in the last section.
The formal integrability of the equation given in the
fourth example is trivial, because we can write down all its symmetries explicitly.
For the other three equations, we have the following proposition:

\begin{proposition}\label{intb-kdv}
The KdV equation (\ref{kdv}), the Camassa-Holm equation (\ref{ch-eq-a}) and the Burgers equation (\ref{burgers-eq})
are formally integrable.
\end{proposition}
\begin{proof}
According to Proposition \ref{mainlem} and the above construction, we know that the flow (\ref{sym-1}) for the
KdV equation must take the following form
\begin{equation}\label{kdv-sym}
u_s=h(u)\,u_x+\sum_{k\ge1}\ee^{2k} \frac1{u_x^{m_k}}\sum_{\lm \in \cp^{m_k+2k+1}}
C_{k,\lambda}\,h^{(D_{k,\lambda})}(u)u_\lm,
\end{equation}
where $m_k$ are integers, $C_{k,\lm}$ are rational numbers and $D_{k,\lm}$ are positive integers.
To prove the theorem, we only need to show that for each  monomial
\begin{equation}\label{kdv-sym-2}
u_x^{-m_k} C_{k,\lambda}\,h^{(D_{k,\lambda})}(u)u_\lm
\end{equation}
that appears in the above expression, either $C_{k,\lambda}=0$
or it can be reduced to the form
$$C_{k,\lambda}\,h^{(D_{k,\lambda})}(u) u_{\lm'},\quad \lm'\in \cp.$$
To this end, let us take $h(u)=u^{D_{k,\lm}}$.
 From the classical theory of the KdV equation, we know that the flow (\ref{kdv-sym}) belongs to the KdV hierarchy
and so its right hand side
is a truncated differential polynomial. So the monomial (\ref{kdv-sym-2}) has the required property
and we proved the the formal integrability
of the KdV equation.

In a similar way, we can prove the formal integrability of the Camassa-Holm equation
and Burgers equation. The only technical point we should note is that the analogue of (\ref{kdv-sym})
for the Camassa-Holm equation
need to be slightly modified.  We omit the details here.
The proposition is proved.
\hfill $\Box$ \end{proof}

The following two examples illustrate a procedure to identify the formally integrable
equations among those which possess certain particular forms.

\begin{example}\label{emp-bgs}
Consider the formally integrable equations among those of the form
\begin{equation}\label{bg-1}
u_t=u u_x+\ee (f_1(u) u_{xx}+f_2(u) u_x^2),\quad f_1(u)\neq 0.
\end{equation}
By using the criterion of formal integrability we easily obtain
\begin{equation}\label{bg-2}
f_1(u)=a\,u+b,\quad f_2(u)=-\frac12 a,
\end{equation}
where $a, b$ are arbitrary constants. When $a=0, b=1$ the above equation is just the well known Burgers equation (\ref{burgers-eq}).
If $a\ne0$, we may assume without loss of generality that $a=1$. The Galilean transformation
$x \mapsto x-b\,t,\ t \mapsto t,\ u \mapsto u-b$ converts (\ref{bg-1}) to the form
\begin{equation}\label{mbg-1}
u_t=u\,u_x+\ee\left(u\,u_{xx}-\frac12 u_x^2\right)
\end{equation}
Let $u=w^2$, the equation (\ref{mbg-1}) becomes
\begin{equation}\label{mbg-2}
w_t=w^2\left(w_x+\ee\,w_{xx}\right)
\end{equation}
This equation is also linearizable under the change of the independent variables and is called C-integrable,
see \cite{c-int} and the equation (3.37) there.

The equation (\ref{mbg-2}) admits a recursion operator
\begin{equation}\label{mbg-ro}
R=\ee\,w\,\px+w+w^2\left(w_x+\ee\,w_{xx}\right)\px^{-1}\frac1{w^2}.
\end{equation}
One can prove that $R$ is a hereditary strong symmetry \cite{FF}. So we obtain a hierarchy of symmetries
of the equation (\ref{mbg-2}) which can be
expressed as
\begin{equation}\label{mbg-hy}
w_{t_0}=-w_x,\ w_{t_1}=0,\ w_{t_{2+k}}=R^k\left(w^2\left(w_x+\ee\,w_{xx}\right)\right)=(k+1)w^{k+2}w_x+\cdots.
\end{equation}
To see that all the right hand sides of these symmetries are differential polynomials of the variables
$w_x, w_{xx},\dots$, we introduce a generating function
$$F=\sum_{k=0}^\infty \frac{w_{t_{k+2}}}{\lambda^{k+2}}=\frac{w_{t_2}}{\lambda^2}+\frac{w_{t_3}}{\lambda^3}+\cdots.$$
It satisfies the following equation
\begin{equation}\label{mbg-rf}
R\,F=\lambda\left(F-\frac{w_{t_2}}{\lambda^2}\right).
\end{equation}
Let $w_{t_k}=w^2\,\px h_k,\ H=\sum_{k\ge2}\frac{h_k}{\lm^k}$, the equation (\ref{mbg-rf}) becomes a linear ODE of $H$, its solution is
$$H=\frac1w\sum_{k=0}^\infty \ee^k \left(A\,\px\right)^k \left(A\frac{w+\ee\,w_x}{\lambda}\right),
\mathrm{\ where\ } A=\sum_{l=1}^\infty\frac{w^l}{\lambda^l}.$$
This implies that all the right hand sides of the flows given in (\ref{mbg-hy})
are homogeneous differential polynomials. By using a similar argument as given in the proof of Proposition \ref{intb-kdv},
one can prove that the equation (\ref{mbg-2}) is integrable.

The integrable hierarchy (\ref{mbg-hy}) has the conversation law
\begin{equation}
\left(-\frac1w\right)_{t_k}=\px\,h_k.
\end{equation}
It defines a reciprocal transformation
\begin{equation}\label{reci-0}
dy=\frac1w dx-\sum_{k\ge2}h_k\,dt_k
\end{equation}
which transform the whole hierarchy to the Burgers hierarchy (up to a rescaling).
For more details on the reciprocal transformation, please refer to the next example.
\end{example}

In the above example, the conditions (\ref{bg-2}) can in fact be derived by requiring the
existence of a function $h(u)$ with $h''(u)\ne 0$ such that the resulting functions $W_k$ are
differential polynomials. Similar situation also occurs in other examples that we computed, such as the
one which will be presented below. Based on these examples, we reformulate the conjecture
of Fokas \cite{Fokas1,Fokas2}
on the existence of infinitely many symmetries of a scalar evolutionary PDE for
the class of equation considered in this paper as follows:

\begin{conjecture}\label{def-sym}
A generalized evolutionary PDE of the form (\ref{zh-0}) is formally integrable
iff there exists a smooth function
$h(u)$ satisfying
\begin{equation}
h''-\frac{h'}{f'} f''\ne 0,
\end{equation}
such that the flow (\ref{sym-1}) that is obtained  From the equation (\ref{sym-2}) by the reducing transformation
of (\ref{zh-0}) gives a symmetry of (\ref{zh-0}), i.e., the functions $W_k$ are homogeneous differential polynomials
of degree $k$.
\end{conjecture}
%%%%%

\begin{example}
Consider the equation of the form
\begin{equation}\label{gkdv}
u_t=u u_x+\ee^2 \left(g_1(u) u_{xxx}+ g_2(u) u_x u_{xx}+g_3(u) u_x^3\right), \quad g_1(u)\ne 0.
\end{equation}
%%%%%%%%%%%%%%%%%
\begin{lemma}
If the equation (\ref{gkdv})
is formally integrable, then the functions $g_1, g_2, g_3$ must satisfy the equations
\begin{eqnarray}
&&9\,g_1^2\,g_2''-6\,g_1\,g_2\, g_1''-9\,g_1\,g_1'\,g_2'\nn\\
&&\quad -18\,g_1\,g_2\,g_2'+8\,g_2\,{g_1'}^2
+12\,g_2^2\,g_1'+4\,g_2^3=0, \label{gkdv-1}\\
&&12\,g_1^2\,g_1'''-8\,g_1\,g_1'\,g_1''-12\,g_1\,g_2\,g_1''\nn\\
&&\quad-3\,g_1\,g_1'\,g_2'+4\,{g_1'}^3+6\,g_2\,{g_1'}^2+
2\,g_2^2\,g_1'=0 \label{gkdv-2}
\end{eqnarray}
and the relation
\begin{equation}\label{gkdv-3}
g_3=\frac1{72\,g_1} \left(6\,g_2^2-30\,g_2\,g_1'-4\,{g_1'}^2+27\,g_1\,g_2'+12\,g_1\,g_1''\right).
\end{equation}
\end{lemma}
\begin{proof}
By using the criterion of integrability, we easily arrive at the result of the lemma  From the
polynomiality property of
the first few $W_k$. The lemma is proved.
\hfill $\Box$ \end{proof}

\begin{lemma}
If the conditions (\ref{gkdv-2}), (\ref{gkdv-3}) hold true and $g_1(u)$ is not a constant, then there exists
a reciprocal transformation
\begin{equation}\label{gkdv-rt-1}
dy=f(u)\,dx+\rho(u,u_x,u_{xx})dt,\ ds=dt,
\end{equation}
such that $u$ satisfies an equation of the following form:
\begin{equation}\label{gkdv-rt}
u_s=\tilde{f}(u)\,u_y+\ee^2\left(u_{yyy}+ \tilde{g}_2(u)\,u_{yy}\,u_y+\tilde{g}_3(u)\,u_y^3\right).
\end{equation}
\end{lemma}
\begin{proof}
To define the reciprocal transformation (\ref{gkdv-rt-1}), $f(u)$ must be the density of
a conversation law, i.e. there
exists a function $\rho(u,u_x,u_{xx})$ such that
$$\left(f(u)\right)_t=\left(\rho(u,u_x,u_{xx})\right)_x.$$
By using the equation (\ref{gkdv}), it's easy to see that $f(u)$ must satisfy the following equation:
\begin{equation}\label{gkdv-rt-2}
g_1\,f'''+2\,g_1'\,f''+g_1''\,f'-g_2\,f''-g_2'\,f'+2\,g_3\,f'=0,
\end{equation}
and the flux $\rho$ is given by
\begin{equation}\label{gkdv-rt-3}
\rho=u\,f-F+\ee^2\left(g_1\,f'\,u_{xx}+\frac12\left(g_2\,f'-g_1\,f''-g_1'\,f'\right)u_x^2\right),
\end{equation}
where $F(u)=\int f(u)\,du$.

Suppose we have a function $f(u)$ satisfying (\ref{gkdv-rt-2}) and a function $\rho$ given by (\ref{gkdv-rt-3}),
then the reciprocal
transformation (\ref{gkdv-rt-1}) is well-defined and it converts the equation (\ref{gkdv}) to the following one
\begin{equation}\label{gkdv-rt-4}
u_s=F(u)\,u_y+\ee^2 \left(g_1(u)\,f(u)^3\,u_{yyy}+ \tilde{g}_2(u)\,u_{yy}\,u_y+\tilde{g}_3(u)\,u_y^3\right).
\end{equation}
To complete the proof, we only need to show that $f=g_1^{\frac13}$ is a solution of the equation (\ref{gkdv-rt-2}).
In fact, after the substitutions of $f=g_1^{\frac13}$ and  (\ref{gkdv-3}) into the equation (\ref{gkdv-rt-2}),
we obtain
an equation which is equivalent to (\ref{gkdv-2}). The lemma is proved.
\hfill $\Box$ \end{proof}

Now let us perform a Miura type transformation of the first kind
$$u\mapsto\tilde{u}=\tilde{f}(u)$$
on the equation (\ref{gkdv-rt}),
we obtain a equation of the following form
\begin{equation}\label{gkdv-4}
{\tilde u}_s={\tilde u}\,{\tilde u}_y+\ee^2\left({\tilde u}_{yyy}+ {\bar g}_2({\tilde u})\,
{\tilde u}_{yy}\,{\tilde u}_y+{\bar g}_3({\tilde u})\,{\tilde u}_y^3\right).
\end{equation}
Since the above reciprocal and Miura type transformations keep the
formal integrability property, the functions ${\bar g}_1=1, {\bar
g}_2, {\bar g}_3$ also satisfy the conditions
(\ref{gkdv-1})-(\ref{gkdv-3}), i.e.
\begin{eqnarray}
&&{\bar g}_2''-2\,{\bar g}_2\,{\bar g}_2'+\frac49\,{\bar g}_2^3=0, \label{gkdv-5}\\
&&{\bar g}_3=\frac1{24}\left(2\,{\bar g}_2^2+9\,{\bar g}_2'\right). \label{gkdv-6}
\end{eqnarray}

\begin{lemma}
Suppose the functions ${\bar g}_2({\tilde u}),{\bar g}_3({\tilde u})$
satisfy the conditions (\ref{gkdv-5}), (\ref{gkdv-6}), then there exists
a Miura type transformation
\begin{equation}
u\mapsto\bar{u}={\tilde u}+\ee\,f_0({\tilde u})\,{\tilde u}_y+\ee^2\left(f_1({\tilde u})\,{\tilde u}_{yy}+f_2({\tilde u})\,{\tilde u}_y^2\right)
\end{equation}
which transforms the equation (\ref{gkdv-4}) into the KdV equation
$$
\bar{u}_s=\bar{u}\,\bar{u}_y+\ee^2\,\bar{u}_{yyy}.
$$
\end{lemma}
\begin{proof}
The ODE (\ref{gkdv-5}) has the following general solution
$${\bar g}_2({\tilde u})=-\frac{3f'({\tilde u})}{2f({\tilde u})},\ f({\tilde u})=a_0+a_1\,{\tilde u}+a_2\,{\tilde u}^2,
$$
where $a_0,a_1,a_2$ are arbitrary constants. Let's take
$$
f_0=\frac{\sqrt{\frac32}\sqrt{4\,a_0\,a_2-a_1^2}}{\sqrt{f}\sqrt{f'+2\sqrt{a_2}\sqrt{f}}},\
f_1=\frac{3\sqrt{a_2}}{2\sqrt{f}},\ f_2=f_1'-\frac16\,f_1^2.
$$
The lemma is proved after a direct verification.
\hfill $\Box$ \end{proof}

The above three Lemmas show that the equation (\ref{gkdv}) is formally integrable if the functions
$g_1, g_2, g_3$ satisfy (\ref{gkdv-1})-(\ref{gkdv-3}), and modulo reciprocal and Miura type transformations
there is only one formally integrable equation, namely, the KdV equation. So our formal integrability
in fact leads to integrable equations in the common sense.
%%%%%%%%%%%%%%%%%

To see some concrete formally integrable equations of the form (\ref{gkdv})
that are in different guises of the KdV equation, let us further impose the additional conditions on
the functions $g_1, g_2, g_3$ by
requiring that the right hand side of (\ref{gkdv}) is graded homogeneous with respect to the following grading
\begin{equation}\label{homo-1}
\deg u=1, \ \deg \px^m u=1-m,\ \deg\ee =1+\dl,
\end{equation}
where $\dl$ is certain constant. Since $g_1(u) \ne 0$, we can take
\begin{equation}\label{homo-2}
g_1(u)=u^{1-2 \dl},\ g_2(u)=\al\,u^{-2 \dl},\quad \al \mbox{ is constant}.
\end{equation}
Then the equations (\ref{gkdv-1}), (\ref{gkdv-2}) have the following solutions $(\al, \dl)$:
\begin{eqnarray}
&&(0, \frac12),\ (-\frac32, \frac12),\ (-3, \frac12),\ (-2, 0),\ (-1,0),
\ (0, -\frac14),\nn\\
&& (-\frac34, -\frac14),\ (0, -\frac12),\ (-6, 2).\nn
\end{eqnarray}
They correspond to the following equations:
\begin{eqnarray}
(0, \frac12) &:& u_t=u\,u_x+\ee^2 u_{xxx}, \label{gkdv-n1} \\
(-\frac32, \frac12) &:& w_t=w^2\,w_x+\ee^2\,w_{xxx}, \ \ w^2=u, \label{gkdv-n2} \\
(-3, \frac12) &:& u_t=u\,u_x+\ee^2\left(u_{xxx}-\frac{3}{u}\,u_x\,u_{xx}+\frac{15}{8\,u^2}\,u_x^3\right),
\label{gkdv-n3} \\
(-2, 0) &:& w_t=w^3\,w_x+\ee^2\,w^3\,w_{xxx}, \ \ w^3=u, \label{gkdv-n4} \\
(-1, 0) &:& w_t=w^3\,w_x+\ee^2\left(3\,w^2\,w_x\,w_{xx}+w^3\,w_{xxx}\right), \ \ w^3=u, \label{gkdv-n5} \\
(0, -\frac14) &:& w_t=w^2\,w_x+\ee^2\left(3\,w^2\,w_x\,w_{xx}+w^3\,w_{xxx}\right),
\ \ w^2=u, \label{gkdv-n6} \\
(-\frac34, -\frac14) &:& w_t=w^2\,w_x+\ee^2\left(\frac32\,w^2\,w_x\,w_{xx}+w^3\,w_{xxx}\right),
 w^2=u, \label{gkdv-n7} \\
(0, -\frac12) &:& u_t=u\,u_x+\ee^2\left(u^2\,u_{xxx}+\frac19 u_x^3\right), \label{gkdv-n8} \\
(-6, 2) &:& w_t=\frac{w_x}{w}+\ee^2\,w^3\,w_{xxx}, \ \ \frac1w=u. \label{gkdv-n9}
\end{eqnarray}

The first two equations (\ref{gkdv-n1}) and (\ref{gkdv-n2}) are the KdV and mKdV equation respectively. The third one (\ref{gkdv-n3})
is equivalent to the KdV equation (\ref{gkdv-n1}) by the following Miura type transformation
\begin{equation}\label{gkdv-miura}
u\mapsto\tilde{u}=u+\ee^2\left(\frac{3\,u_{xx}}{2\,u}-\frac{15\,u_x^2}{8\,u^2}\right),
\end{equation}
where $u$ satisfies (\ref{gkdv-n3}) and $\tilde{u}$ satisfies (\ref{gkdv-n1}).

The equations (\ref{gkdv-n1})-(\ref{gkdv-n3}) have the following pairs of conserved quantities respectively:
\begin{eqnarray*}
\int u\,dx,\ \int u^2\,dx; \
\int w\,dx,\ \int w^2\,dx;\
\int \sqrt{u}\,dx,\ \int \frac1{\sqrt{u}}\,dx.
\end{eqnarray*}
By using these conserved quantities we can define six reciprocal transformations, they relate the remaining six equations
(\ref{gkdv-n4})-(\ref{gkdv-n9}) to the first three equations (\ref{gkdv-n1})-(\ref{gkdv-n3}).
Let us consider in detail the reciprocal transformation that is defined by the second conserved quantity
$\int u^2(x,t) dx $ of the equation (\ref{gkdv-n1}). The conservation law is given by
\begin{equation}\label{recip-5}
\left(u^2\right)_t=\rho_x,\ \mbox{where } \rho=\frac23\,u^3+\ee^2\left(2\,u\,u_{xx}-u_x^2\right).
\end{equation}
Thus we can define the following
reciprocal transformation:
\begin{equation}\label{reci-1}dy=u^2\,dx+\rho\,dt,\ ds=dt,\end{equation}
its inverse is given by
\begin{equation}\label{reci-2}
dx=\frac1{u^2}\,dy-\left(\frac23\,u+\ee^2\left(\frac{2\,u_{xx}}{u}-\frac{u_x^2}{u^2}\right)\right)\,ds,\quad dt=ds.
\end{equation}
 From the definition (\ref{reci-1}) we obtain $\px=u^2\,\pp_y$, so
\begin{equation}\label{recip-4}
u_x=u^2\,u_y,\ u_{xx}=u^2\left(u^2\,u_{yy}+2\,u\,u_y^2\right).
\end{equation}
Now the reciprocal transformation (\ref{reci-2}) can be rewrite as
\begin{equation}\label{reci-3}
dx=\frac1{u^2}\,dy-\left(\frac23\,u+\ee^2\left(2\,u^3\,u_{yy}+3\,u^2\,u_y^2\right)\right)\,ds,\quad dt=ds.
\end{equation}
So in terms of the new independent variables $y, s$, the function $u$ satisfies the following equation:
\begin{equation}\label{recip-6}
u_s=\frac{u^3}3\,u_y+\ee^2\left(u^6\,u_{yyy}+6\,u^5\,u_y\,u_{yy}+3\,u^4\,u_y^3\right).
\end{equation}
By performing the rescaling $s\mapsto\tilde{s}=s/3,\ee\mapsto\tilde{\ee}=\sqrt{3}\,\ee$ and the Miura transformation of first
kind $u\mapsto\tilde{u}=u^3$, the above equation is transformed to
\begin{equation}
\tilde{u}_{\tilde{s}}=\tilde{u}\,\tilde{u}_y+\tilde{\ee}^2\left(\tilde{u}^2\,\tilde{u}_{yyy}
+\frac19\tilde{u}_y^3\right),
\end{equation}
which coincides with the equation (\ref{gkdv-n8}).

For the other cases, we only point out the following facts:
\begin{enumerate}
\item{}The equation (\ref{gkdv-n1}) is transformed to the equation (\ref{gkdv-n6}) by the reciprocal transformation
defined by the conserved quantity $\int u\,dx$.
\item{}
The equation (\ref{gkdv-n2}) is transformed to the equations (\ref{gkdv-n5}), (\ref{gkdv-n7}) respectively by the
reciprocal transformations defined by conserved quantities $\int w\,dx$ and $\int w^2\,dx$.
\item{}
The equation (\ref{gkdv-n3}) is transformed to the equations (\ref{gkdv-n4}), (\ref{gkdv-n9}) respectively by the
reciprocal transformations defined by conserved quantities $\int \sqrt{u}\,dx$ and $\int 1/\sqrt{u}\,dx$.
\end{enumerate}
Note that the equation (\ref{gkdv-n5}) and (\ref{gkdv-n6}) are two cases of the S-integrable equation (3.55) given in \cite{c-int}
\end{example}

%%%%%

The analysis of the above two examples shows that the formal integrability condition is
rather rigid. Modulo the Miura type transformations and reciprocal transformations,
the Burgers equation and the KdV equation are the
only integrable equations among the two classes of equations of the form (\ref{bg-1}) and (\ref{gkdv}).
The following conjecture generalize the results of the first example:

\begin{conjecture}
All the formally integrable equations written in the reduced form
\begin{eqnarray}\label{conj-2}
&&u_t=u\,u_x+\ee\,(f_2(u)\,u_2+f_{11}(u)u_1^2)+\ee^2 f_3(u)u_3\nn\\
&&\qquad +
\ee^3(f_4(u)\,u_4+f_{22}(u)u_2^2)+\cdots
\end{eqnarray}
with $f_2(u)\ne0$ are parameterized by the functions $f_2(u)$ and $f_{11}(u)$. In other words, the formal
integrability
condition for the
equation (\ref{conj-2}) with $f_2(u)\ne0$ uniquely determines  all the
coefficient functions $f_{\lambda}(u)$ with $|\lambda|>2$
through the functions  $f_2(u)$ and $f_{11}(u)$,  for example
\begin{eqnarray*}
f_3&=&\frac43\,f_2\,f_{11}+\frac23\,f_2\,f_2',\\
f_4&=&\frac53\,f_2^2\,f_{11}'+\frac13\,f_{2}^2\,f_{2}''+2\,f_{2}\,f_{11}^2+
\frac13\,f_2\,(f_2')^2+\frac53\,f_{11}\,f_2\,f_2',\\
f_{22}&=&-\frac{10}3\,f_{11}^3-f_{11}^2\,f_2'+\frac13\,f_{11}\,(f_2')^2-\frac{29}3\,f_2\,f_{11}\,f_{11}'
-\frac{10}3\,f_2\,f_{11}\,f_2'' \\
&&\quad-4\,f_2\,f_2'\,f_{11}'-f_2\,f_2'\,f_2''-3\,f_2^2\,f_{11}''-f_2^2f_2''',\ \cdots
\end{eqnarray*}
Moreover, modulo certain reciprocal transformation a formal integrable equation of the above form
is equivalent, under a Miura type
transformation, to a formal symmetry of the Burgers equation (\ref{burgers-eq}).
\end{conjecture}

The parameter space for the class of integrable equations of the form (\ref{conj-2}) with $f_2(u)=0$ is unknown yet.
It seems that there exist infinitely many function
parameters in this space. We hope that the additional condition of possessing a hamiltonian structure will
restrict this parameter space to a manageable one. We will consider
this problem in subsequent publications.

\section{Conclusion}
We have proved the quasitriviality of a generalized evolutionary PDE of the form (\ref{zh-0}) and proposed
a criterion for its integrability. Due to Proposition \ref{mainlem}, the proof of Theorem \ref{zh-thm1}
gives a constructive algorithm to obtain an explicit expression of the quasi-Miura transformation that reduces the
equation (\ref{zh-0}) to its leading term equation (\ref{zh-05-10}). Also due to Proposition \ref{mainlem},
the explicit expression of the flows of the form (\ref{sym-1}) that commute with (\ref{zh-0})
can be obtained in a direct way as explained in the
beginning of Sec.\,\ref{sec-5}, thus the algorithm of testing the formal integrability of an equation of the form (\ref{zh-0})
can be encoded into a simple computer program.

Now it is natural to ask the
question whether the quasitriviality property can be generalized to systems of generalized evolutionary PDEs of the form
\begin{equation}\label{mc}
u^i_t=\sum_{j=1}^n  \lm_j^i({\bf u}) u^j_x+\sum_{k\ge 1} \ee^k Q^i_{k}({\bf u},{\bf u}_x,\dots,{\bf u}^{(k+1)}),
\quad i=1,\dots,n\ge 2,
\end{equation}
where ${\bf u}=(u^1,\dots, u^n)$, $\lm^i_j$ are smooth functions of ${\bf u}$ and $Q^i_k$ are
homogeneous polynomials of $\px^m u^l,\ l=1,\dots, n,\ m=1,\dots,k+1$ with degree $k+1$, here
we define $\deg \px^m u^l=m$. It is proved in \cite{DLZ} that if the above system has a semisimple
bihamiltonian structure then it is quasitrivial, i.e., there exists a quasi-Miura transformation of the form
\begin{equation}
u^i\mapsto {v}^i=u^i+\sum_{k\ge 1} \ee^k F^i_k({\bf u},{\bf u}_x,\dots),\quad i=1,\dots, n
\end{equation}
that reduces the above system to a system given by its leading terms
\begin{equation}
{v}^i_t=\sum_{j=1}^n  \lm_j^i({\bf {v}}) {v}^j_x .
\end{equation}
We expect that the condition of possessing
a semisimple bihamiltonian structure could be replaced by a much weaker one in order to ensure the quasitriviality
of the systems of the above form.

The definition of formal integrability for equations of the form (\ref{zh-0}) can be directly generalized
to systems of equations of the form (\ref{mc}), we plan to study the problem of its quasitriviality and integrability in subsequent
publications.

\vskip 0.4truecm \noindent{\bf Acknowledgments.} The authors are
grateful to Boris Dubrovin for encouragements and helpful discussions.
The researches of Y.Z. were partially
supported by the Chinese National Science Fund for Distinguished
Young Scholars grant No.10025101 and the Special Funds of Chinese
Major Basic Research Project ``Nonlinear Sciences''.

\end{document}